\documentclass{article}

\usepackage{amsmath,amssymb,amsfonts}
\usepackage{a4wide}
\usepackage{color}
\usepackage{bm}
\usepackage{graphicx}

\DeclareMathOperator*{\sgn}{sgn}
\DeclareMathOperator*{\End}{End}

\author{Chihiro Matsui\\[3ex]
{\it Department of Mathematical Informatics, The University of Tokyo} \\
{\it 7-3-1 Hongo, Bunkyo-ku, Tokyo 113-8656, Japan}
}
\date{\today}
\title{Spinon excitations in the spin-$1$ XXZ chain and hidden supersymmetry}

\begin{document}
\maketitle

\begin{center}
{\bf Abstract}
\end{center}
\bigskip
{\small
We study spinon excitations of the integrable spin-$1$ (Fateev-Zamolodchikov; FZ) chain and their relation to the hidden supersymmetry. Using the notion of the supercharges earlier introduced to the spin chains, which change the system length by one, we found that they nontrivially act on one of two kinds of the degrees of freedom for the FZ chain. Their actions were obtained to be the same as those of the supercharges defined on the supersymmetric sine-Gordon model, the low-energy effective field theory of the FZ chain. Moreover, we construct the eigenstates which are invariant under the supersymmetric Hamiltonian given as the anti-commutator of the supercharges. 
}

\section{Introduction} \label{sec:intro}
The integrable spin chains have been studied since a long ago and various methods have been developed to diagonal their Hamiltonians. One of the most successful methods for the finite XXZ spin chains is the Bethe ansatz method~\cite{bib:B31}, which allows us not only to derive eigenfunctions but also to calculate correlation functions~\cite{bib:DM09, bib:GKS04, bib:K01, bib:KMT00, bib:KBI93, bib:S89} through the quantum inverse scattering method~\cite{bib:TF79}. Another successful method is the $q$-vertex operator approach~\cite{bib:BW94, bib:I93, bib:JMMN92, bib:JM95, bib:K94}, which is valid for infinite chains. 
Both methods are based on the common property of the integrable spin chains, the factorizability of the scattering $S$-matrix resulting from the Yang-Baxter equation. For the XXZ spin chain with arbitrary spin $\frac{\ell}{2}$, the $S$-matrix is known to be decomposed into the $SU(2)$ part and the restricted solid-on-solid (RSOS) part~\cite{bib:R91}. 
In the Bethe ansatz method, an excitation particle is a magnon, which is defined as a flipped spin in the vacuum given by fully polarized spins. Each magnon carries rapidity, which takes complex number, and magnons carrying rapidities with the same real part but with the different imaginary parts make a bound state. Such a set of rapidities is called string solutions~\cite{bib:T82}. 
On the other hand, in the $q$-vertex operator approach, an excitation particle is a spinon, which emerges at an unpaired bond in the vacuum given by singlet paired spins. Each spinon carries spin $\frac{1}{2}$~\cite{bib:FT81} and this freedom shows up in the $SU(2)$ part of the $S$-matrix. 

The spin-$\frac{1}{2}$ XXZ chain reveals the quantum phase transition depending on the anisotropy~\cite{bib:T99}. In the massless regime, its low energy excitations are described by the quantum sine-Gordon model. For the spin-$1$ case, the low energy effective field theory is the supersymmetric sine-Gordon model~\cite{bib:IO93} and for the arbitrary spin-$\frac{\ell_{\geq 3}}{2}$ case, it is expected that the low energy excitations are described by the fractional supersymmetric sine-Gordon model through the comparison of the structure of the $S$-matrices~\cite{bib:D03}. 
What is interesting is, although the supersymmetry is not explicitly obtained in the spin chains, their effective field theory possess the supersymmetry. Reflecting the structure of the $S$-matrix, the supersymmetric effective field theory has an asymptotic state labeled by a soliton charge and a pair of RSOS indices. The generators of the superalgebra non-trivially act on the asymptotic state by changing the RSOS indices but by leaving the soliton charge preserved~\cite{bib:BPT02, bib:HM97, bib:S90, bib:Z89}. This motivates us to consider how the superalgebra acts on a spinon state of the spin chain, which also possesses a spinon spin and a pair of RSOS indices. 

Recently, the discrete analog of the supersymmetry was introduced to the spin chains~\cite{bib:H13, bib:HF12, bib:MM14, bib:YF04}. The idea originates in the fermion representation of the spin operator $\vec{S}_j = f_j^{\dag} \vec{\sigma}_{\alpha}^{\beta} f_{j\beta}$, which assign a fermion to each lattice site. Since the supersymmetry exchanges a bosonic state and a fermionic state, {\it i.e.} it changes the fermion number by one, the supercharges were defined to change the system length by one. Subsequently, the action of the supercharge was given on the spin basis as a map $V \mapsto V \otimes V$ or $V \otimes V \mapsto V$. 
Although it is hard to deal with such an operator that changes the system size in the spin basis, we expect that the action of the superalgebra is clearly understood on the RSOS part of the spinon basis. Indeed, the $\ell = 2$ case possesses the $\mathcal{N} = 1$ supersymmetry and the total magnetization is conserved, which implies that the spinon number could be conserved as well. 

Therefore, the aim of this paper is to investigate the discrete analog of supersymmetry on the spinon basis. 
For simplicity, we limit our interest to the $\mathcal{N} = 1$ case, in which the total magnetization is a conserved quantity of the superalgebra. The $\mathcal{N} = 1$ supersymmetry is observed only for the spin-$1$ case, and therefore we focus on the Fateev-Zamolodchikov (FZ) spin chain known as the spin-$1$ integrable XXZ chain~\cite{bib:ZF80}. Moreover, we restrict our discussion to the infinite spin chain, where the whole system is $U_q(\widehat{sl}_2)$-invariant and the spinon excitations are well-described by the vertex operator. Since the supersymmetry was discussed only for the finite systems either with the periodic boundary or open boundaries~\cite{bib:H13, bib:HF12, bib:YF04}, we modify it in compatible with the infinite spin chain. 

This paper is organized as follows. In the next section, we review the known properties of the ZF chain and its spinon basis introduced in \cite{bib:NY96}. Section~\ref{sec:SUSY} is devoted to the attempt to combine the notion of spinon basis with the discrete analog of supersymmetry, by discussing the actions of superalgebra on the spinon basis. For better understanding, we introduce the graphical representation of the restricted paths. At the same time, we construct the superspace where all the eigenvectors except for the ground state make superpartners. The relation to Haldane's spinon motifs~\cite{bib:HHTBP92} is also pointed out. The conclusion and future works are given in the last section.

\section{Spinon excitations of the spin-$1$ XXZ chain} \label{sec:spinon}
The Fateev-Zamolodchikov (FZ) spin chain is known as the integrable spin-$1$ XXZ model. The integrability lies in the factorizable bulk scattering matrix obtained as the trigonometric solution (a nine-by-nine matrix) of the Yang-Baxter equation~\cite{bib:ZF80}. The Hamiltonian of the FZ spin chain is written in the following form~\cite{bib:H13}: 
\begin{equation} \label{eq:H_spin1}
 H = \sum_{j \in \mathbb{Z}} h_{j,j+1}
  = \sum_{j \in \mathbb{Z}} 
  \left(
   \sum_{a,b=1}^3 A_{ab} S_j^a S_j^b S_{j+1}^a S_{j+1}^b
   - \sum_{a=1}^3 J_a(S_j^a S_{j+1}^a + 2(S_j^a)^2)
  \right), 
\end{equation}
where the coupling constants are chosen as 
\begin{equation} \label{eq:coupling_const}
\begin{split}
 &J_1 = A_{11} = J_2 = A_{22} = 1, \quad J_3 = A_{33} = \dfrac{(q - q^{-1})^2}{2}\\
 &A_{12} = A_{21} = 1, \quad A_{13} = A_{31} = A_{23} = A_{32} = q - 1 + q^{-1}. 
\end{split}
\end{equation}
Throughout this paper, we consider only the infinite chain and therefore the index $j$ in \eqref{eq:H_spin1} runs over all integers. The spin operator $S_j^a$ is a spin-$1$ operator which nontrivially acts on the $j$th site: 
\begin{equation}
 S_j^a := 1 \otimes \cdots \otimes 1 \otimes \underbrace{S^a}_j \otimes \cdots \otimes 1. 
\end{equation}
By choosing the basis as 
\begin{equation} \label{eq:basis_spin}
 |0\rangle = \begin{pmatrix} 1 \\ 0 \\ 0 \end{pmatrix}, \quad
 |1\rangle = \begin{pmatrix} 0 \\ 1 \\ 0 \end{pmatrix}, \quad
 |2\rangle = \begin{pmatrix} 0 \\ 0 \\ 1 \end{pmatrix}, 
\end{equation}
the local actions of the spin operators are written as follows: 
\begin{equation} \label{eq:def_spin}
 S^1 = \frac{1}{\sqrt{2}}
  \begin{pmatrix}
   0 & 1 & 0 \\ 1 & 0 & 1 \\ 0 & 1 & 0
  \end{pmatrix}, 
  \quad
  S^2 = \frac{1}{\sqrt{2}}
  \begin{pmatrix}
   0 & -i & 0 \\ i & 0 & -i \\ 0 & i & 0
  \end{pmatrix},
  \quad
  S^3 = 
  \begin{pmatrix}
   1 & 0 & 0 \\ 0 & 0 & 0 \\ 0 & 0 & -1
  \end{pmatrix}. 
\end{equation}
Thus, we have $S^3 |p(l)\rangle = (-p(l)+1) |p(l)\rangle$ for $p(l) \in \{0,1,2\}$. We write a tensor product state by 
\begin{equation} \label{eq:tensor_p}
 \otimes_{l \in \mathbb{Z}} |p(l)\rangle := |\dots, p(l-1), p(l), p(l+1), \dots \rangle. 
\end{equation}
Here $l$ indicates the $l$th site of the spin chain. A basis vector of the whole system is written as a linear combination of tensor product states. 

\subsection{$U'_q(\widehat{sl}_2)$-invariance of the FZ model}
In the case of the infinite spin chain, the choice of the coupling constants \eqref{eq:coupling_const} makes the Hamiltonian \eqref{eq:H_spin1} $U_q(\widehat{sl}_2)$-invariant. Let $V(\lambda_m)$ the three dimensional irreducible highest weight $U_q(\widehat{sl}_2)$-module with the highest weight $\lambda_m$. Then the action of $U_q'(\widehat{sl}_2)$ on $V^{(2)}$ ($\pi: U_q'(\widehat{sl}_2) \to \End(V^{(2)})$) on the basis \eqref{eq:basis_spin} is given by 
\begin{equation} \label{eq:def_uq}
\begin{split}
 &\pi(e^{(0)}) = \pi(f^{(1)}) = 
 \begin{pmatrix} 0 & 0 & 0  \\ 1 & 0 & 0 \\ 0 & \frac{q+q^{-1}}{2} & 0 \end{pmatrix}, \\
 &\pi(e^{(1)}) = \pi(f^{(0)}) = 
 \begin{pmatrix} 0 & \frac{q+q^{-1}}{2} & 0 \\ 0 & 0 & 1 \\ 0& 0 & 0 \end{pmatrix}, \\
 &\pi(k^{(1)}) = \pi((k^{(0)})^{-1}) = 
 \begin{pmatrix} q^2 & 0 & 0 \\ 0 & 1 & 0 \\ 0 & 0 & q^{-2} \end{pmatrix}. 
\end{split}
\end{equation}
They have the comultiplication given by 
\begin{equation} \label{eq:comulti}
\begin{split}
 &\Delta^{(\infty)}(e^{(i)}) = 
  \sum_{j \in \mathbb{Z}} \cdots \otimes k^{(i)} \otimes \underbrace{e^{(i)}}_j \otimes 1 \otimes \cdots, \\
 &\Delta^{(\infty)}(f^{(i)}) = 
  \sum_{j \in \mathbb{Z}} \cdots \otimes 1 \otimes \underbrace{f^{(i)}}_j \otimes (k^{(i)})^{-1} \otimes \cdots, \\
 &\Delta^{(\infty)}(k^{(i)}) = \cdots \otimes k^{(i)} \otimes k^{(i)} \otimes k^{(i)} \otimes \cdots. 
\end{split}
\end{equation}
From \eqref{eq:def_uq} and \eqref{eq:comulti}, one can check the $U'_q(\widehat{sl}_2)$-invariance of the Hamiltonian $[H,\,U'_q(\widehat{sl}_2)] = 0$.

\subsection{Path realization at $q \to 0$}
In the $q \to 0$ limit, the properly normalized Hamiltonian consists only of the operator $S^3$: 
\begin{equation} \label{eq:Hamiltonian_q0}
 q^2 H \underset{q \to 0}{\longrightarrow} 
  H_0 = \frac{1}{2}\sum_{j \in \mathbb{Z}} 
  \left(
   (S_j^3 S_{j+1}^3)^2 - S_j^3 S_{j+1}^3
  \right) 
\end{equation}
and thus becomes diagonal in the tensor products of \eqref{eq:basis_spin}. Therefore, the eigenvectors are written in a form of \eqref{eq:tensor_p}. These eigenvectors are interpreted as paths $(p(l))_{l \in \mathbb{Z}}$~\cite{bib:NY96}. 

There are three degenerate ground state with the vector elements given by 
\begin{equation}
 \bar{p}_m(l) := (p(l))_{l \in \mathbb{Z}}\, |\, 
  p(l) = m + (2 - 2m) \varepsilon(l) \quad m =0,1,2. 
\end{equation}
$\varepsilon(l)$ takes $0$ for even $l$, while $1$ for odd $l$. A sequence $\bar{p}_m(l)$ is called a ground state path. Under the fixed boundary conditions such that 
\begin{equation}
 p(l) = 
    \begin{cases}
   \bar{p}_m(l) &  l \gg 0, \\
   \bar{p}_{m'}(l) & l \ll 0,
    \end{cases}
\end{equation}
the lowest energy state is realized by the vector elements given by 
\begin{equation} 
 \bar{p}_{m,m'}(l) := (p(l))_{l \in \mathbb{Z}}\, \left|\, 
  \begin{matrix}
   p(l) = \bar{p}_m(l) &  l (\in \mathbb{Z}) > 0 \\
   p(l) = \bar{p}_{m'}(l) & l (\in \mathbb{Z}) \leq 0 
  \end{matrix}
  \right..
\end{equation}
This sequence is called an $(m,m')$ ground state path. We simply call an $(m,m')$ path if a sequence satisfies 
\begin{equation}
 p(l) = \bar{p}_{m,m'} \quad |l| \gg 0. 
\end{equation}
There exists a bijection between an $(m,m')$ path and the associated crystal $B(\lambda_m)$ to the crystal base of the level $2$ irreducible highest weight $U_q(\widehat{sl}_2)$-module with the highest weight $\lambda_m$~\cite{bib:NY96}: 
\begin{equation} \label{eq:crystal2path}
 B^{(2)}(\lambda_m) \otimes B^{(2)}(\lambda_{m'})^* = 
  \{p = (p(l))_{l \in \mathbb{Z}}\, |\, p(l) \in \{0,1,2\},\, p(l) = \bar{p}_{m,m'}\, (|l| \gg 0)\}. 
\end{equation}
Here we denote the associated crystal to the crystal base of $\ell+1$ dimensional $U_q(\widehat{sl}_2)$-module by $B^{(\ell)}$. 

%Given an $(m,m')$ ground state path $\bar{p}_{m,m'}$ whose component is given by 
%\begin{equation} 
% \bar{p}_{m,m'}(l) = 
%  \begin{cases}
%   \bar{p}_m(l) := m + (2 - 2m) \varepsilon(l) &  l (\in \mathbb{Z}) > 0, \\
%   \bar{p}_{m'}(l) := m' + (2 - 2m') \varepsilon(l) & l (\in \mathbb{Z}) \leq 0, 
%  \end{cases}
%\end{equation}
%the elements of a single tensor product state \eqref{eq:tensor_p} directly emerge as the components of an $(m,m')$ path ($m,m'\in\{0,1,2\}$): 
%\begin{equation}
% p = (p(l))_{l \in \mathbb{Z}} \,|\,
%  p(l) \in \{0,1,2\},\,
%  p(l) = \bar{p}_{m,m'}\, (|l| \gg 0)
%\end{equation}
%under the proper choice of boundary conditions. 

An $(m,m')$ path admits the domain wall description. If a path contains $n$ domain walls at $l_1, \dots, l_n \in \mathbb{Z}$ ($l_n \geq \cdots \geq l_1$), the path consists of the fragments of $n+1$ ground state paths: 
\begin{equation}
 p(l) = \bar{p}_{m_r}(l), \quad l_{r+1} \geq l > l_r
  %(m_n, \dots, m_0) \in \{0,1,2\}^{n+1}. 
\end{equation}
The adjacent domains must satisfy the adjacency condition: 
\begin{equation} \label{eq:ad_cond}
 |m_j - m_{j-1}| = 1 
\end{equation}
besides the boundary conditions $m_n = m$, $m_0 = m'$. 
For a fixed sequence $(m_n, \dots, m_0)$, a set of domain walls $\{l_1,\dots,l_n\}$ uniquely determines the path components. Thus, we write the $(m_n,m_0)$ path encoded by a fixed sequence $(m_n, \dots, m_0)$ as $[[l_n,\dots,l_1]]_{m_n,m_0}$. The state without domain walls (empty states) are represented by $[[\,]]_{m,m}$. There are three empty states for the choice of $m \in \{0,1,2\}$, corresponding to the three degenerate ground states. 
%If $s'$ domain walls locate at the same position $l_{s+s'} > l_{s+s'-1} = \cdots = l_s > l_{s-1}$, we have $|m_{s+s'-1} - m_s| = s'$. 
Note that, in an $(m,m')$ path, at least $|m - m'|$ domain walls exit, {\it i.e.} the $(m,m')$ ground state path $\bar{p}_{m,m'}$ contains $|m - m'|$ domain walls. 
The bijection \eqref{eq:crystal2path} indicates that there exists the following isomorphism: 
\begin{equation} \label{eq:crystal2DW}
  \bigsqcup_{n=0}^{\infty} \bigsqcup_{l_1,\dots,l_n} [[l_n,\dots,l_1]]_{m,m'}
  \simeq B^{(2)}(\lambda_m) \otimes B^{(2)}(\lambda_{m'})^*. 
\end{equation}
%Especially, we denote the $(m,m)$ ground state path which does not contain domain walls by $[[\,]]$. 
%Here we choose the empty state without domain walls expressed by $[[\,]]$ as the $(0,0)$ groundstate path, {\it i.e.} $[[\,]] = \bar{p}_{0,0}$. 

Let us consider the path consisting of $n$ domain walls at $l_1,\dots,l_n$. In the $q \to 0$ limit, a domain wall $l_j$ is identified with a spinon which carries a pair of RSOS indices and a spinon spin. The RSOS indices are encoded by the ground state paths on both sides of the domain wall $m_{j-1}$, $m_j$, while the spinon spin is defined by 
\begin{equation}
 s_j = -\frac{1}{2}\sgn (p(l_j) + p(l_j+1) - 2). 
\end{equation}
This definition is consistent with the fact that each spinon carries spin $\frac{1}{2}$. In the $q \to 0$ limit, the comultiplication \eqref{eq:comulti} is modified according to \cite{bib:K91}. By denoting the actions of $e^{(i)}$ and $f^{(i)}$ at $q \to 0$ by $\tilde{e}^{(i)}$ and $\tilde{f}^{(i)}$, we consider the following quantities: 
\begin{equation}
 \begin{split}
  &e^{(i)}_{\max} := \max \{k \geq 0; (\tilde{e}^{(i)})^k b \neq 0\}, \quad b \in B^{(2)} \\
  &f^{(i)}_{\max} := \max \{k \geq 0; (\tilde{f}^{(i)})^k b \neq 0\}, \quad b \in B^{(2)}. 
 \end{split}
\end{equation}
Then, $\tilde{e}^{(i)}$ is compatible with the comultiplication defined in \cite{bib:K91} if we choose the action of $\tilde{e}^{(i)}$ as follows: 
\begin{equation} \label{eq:q0comulti_e}
 \Delta^{(\infty)}(\tilde{e}^{(i)}) (\cdots \otimes b_{j+1} \otimes b_j \otimes b_{j-1} \otimes \cdots) =
  \cdots \otimes b_{j+1} \otimes \tilde{e}^{(i)} b_j \otimes b_{j-1} \otimes \cdots 
\end{equation}
for the largest $j$ such that $f^{(i)}_{\max}(b_{j-1}) < e^{(i)}_{\rm max}(b_j)$. 
On the other hand, $\tilde{f}^{(i)}$ is consistently with the definition of the comultiplication by choosing the action of $\tilde{f}^{(i)}$ as 
\begin{equation} \label{eq:q0comulti_f}
 \Delta^{(\infty)}(\tilde{f}^{(i)}) (\cdots \otimes b_{j+1} \otimes b_j \otimes b_{j-1} \otimes \cdots) =
  \cdots \otimes b_{j+1} \otimes \tilde{f}^{(i)} b_j \otimes b_{j-1} \otimes \cdots 
\end{equation}
for the smallest $j$ such that $f^{(i)}_{\max}(b_j) > e^{(i)}_{\max}(b_{j+1})$. 
From these comultiplications, it is easily obtained that $\tilde{e}^{(i)}$ and $\tilde{f}^{(i)}$ simply shift a domain wall: 
\begin{equation}
 \tilde{e}^{(i)} [[l_n, \dots, l_1]] = [[l_n, \dots, l_j+1, \dots, l_1]]
\end{equation}
for the largest $j$ such that $s_j = \frac{1}{2}(-1)^i$, while
\begin{equation}
 \tilde{f}^{(i)} [[l_n, \dots, l_1]] = [[l_n, \dots, l_j-1, \dots, l_1]]
\end{equation}
for the smallest $j$ such that $s_j = \frac{1}{2}(-1)^{i+1}$. If no such $s_j$ exists that satisfies the above condition, we have 
\begin{equation}
 \tilde{x}^{(i)} [[l_n, \dots, l_1]] = 0, \quad
  \tilde{x}^{(i)} = \tilde{e}^{(i)}, \tilde{f}^{(i)}. 
\end{equation}

The adjacency condition \eqref{eq:ad_cond} is expressed by using $p_j \in \{0,1\}$ as 
\begin{equation} \label{eq:ad_cond2}
 m_j - m_{j-1} = (-1)^{p_j}. 
\end{equation}
A sequence $(p_n,\dots,p_1) \in \{0,1\}^n$ is called a level $2$ restricted path. The restricted path associated with an $(m,m')$ path is characterized by 
\begin{equation} \label{eq:res_path}
\begin{split}
 &(-1)^{p_1} + \cdots + (-1)^{p_s} \in \{-m',-m'+1,-m'+2\} \quad 0 \leq s \leq n, \\
 &(-1)^{p_1} + \cdots + (-1)^{p_n} = m. 
\end{split}
\end{equation}
If we set a function $H(p_{s+1},p_s)$ for the components of a restricted paths as 
\begin{equation} \label{eq:energy_func}
 H(1,0) = 1, \quad H(0,0) = H(0,1) = H(1,1) = 0, 
\end{equation}
the energy of the system is given by $\sum_{s=1}^{n-1} H(p_{s+1},p_s)$.

\subsection{Spinon basis}
The spinon creation operator $\varphi_{j}^{*p}$ ($j \in \mathbb{Z}$, $p =0,1$) was introduced as the operator which obeys the following algebraic relations~\cite{bib:DFJMN93, bib:NY96}: 
\begin{eqnarray}
 &\varphi_{j_1}^{*p_1} \varphi_{j_2}^{*p_2} + \varphi_{j_2}^{*p_1} \varphi_{j_1}^{*p_2} = 0, \quad j_1 = j_2\,{\rm mod}\, 2 \land (p_1,p_2) = (1,0), \\
 &\varphi_{j_1}^{*p_1} \varphi_{j_2}^{*p_2} + \varphi_{j_2+1}^{*p_1} \varphi_{j_1-1}^{*p_2} = 0, \quad j_1 \neq j_2\,{\rm mod}\, 2 \land (p_1,p_2) = (1,0), \\
 &\varphi_{j_1}^{*p_1} \varphi_{j_2}^{*p_2} + \varphi_{j_2-2}^{*p_1} \varphi_{j_1+2}^{*p_2} = 0, \quad j_1 = j_2\,{\rm mod}\, 2 \land (p_1,p_2) \neq (1,0), \\
 &\varphi_{j_1}^{*p_1} \varphi_{j_2}^{*p_2} + \varphi_{j_2-1}^{*p_1} \varphi_{j_1+1}^{*p_2} = 0, \quad j_1 \neq j_2\,{\rm mod}\, 2 \land (p_1,p_2) \neq (1,0). 
\end{eqnarray}
These relations were introduced through the inspiration by the relations for the vertex operators~\cite{bib:DFJMN93}. The subscripts $j$ relate to the spinon spins in such a way that even $j$ indicates a $+\frac{1}{2}$ spinon, while odd $j$ indicates a $-\frac{1}{2}$ spinon. 
For these spinon creation operators, the empty state $[[\,]]_{m,m}$ ($m \in \{0,1,2\}$) serves as the Fock vacuum: 
\begin{equation} \label{eq:fock_spinon}
 \varphi_{j_n}^{*p_n} \cdots \varphi_{j_1}^{*p_1} [[\,]]_{m,m}
  = [[j_n-p_n, \dots, j_1-p_1]]_{m,m}. 
\end{equation}
That is, $\varphi_{j_r}^{*p_r}$ creates a domain wall at $j_r-p_r$. 
Since the subscript represents the spinon spin depending on whether it is an even integer or an odd integer, the superscript $p_r$ can be identified with the component of a restricted path defined in \eqref{eq:ad_cond2}. 
Since the domain-wall positions in \eqref{eq:fock_spinon} are set in a non-decreasing order $j_n-p_n \geq \cdots \geq j_1-p_1$, the following condition must be satisfied: 
\begin{equation}
 j_n - 2\sum_{r=1}^{j-1} H(p_{r+1},p_r) \geq \cdots \geq j_2 - 2 H(p_2,p_1) \geq j_1, 
\end{equation}
where $H(p_{r+1},p_r)$ is the energy function defined by \eqref{eq:energy_func}. 
Then, corresponding to \eqref{eq:crystal2DW}, we have a bijection given by 
\begin{equation} \label{eq:crystal2spinon}
 \Big\{ \varphi_{j_n}^{*p_n} \cdots \varphi_{j_1}^{*p_1} \in 
 \bigsqcup_{n=0}^{\infty} \bigsqcup_{j_1-p_1,\dots,j_n-p_n} 
 [[j_n-p_n, \dots, j_1-p_1]]_{m,m} \Big\}
 = B^{(2)}(\lambda_m) \otimes B^{(2)}(\lambda_m)^*. 
\end{equation}

We rewrite \eqref{eq:fock_spinon} as 
\begin{equation} \label{eq:spinon_state}
  \varphi_{2j_{l_n} + i_n}^{*p_n} \cdots \varphi_{2j_{l_1} + i_n}^{*p_1} [[\,]]_{m,m}
  = [[2j_{l_n}+i_n-p_n, \dots, 2j_{l_1}+i_n-p_1]]_{m,m}, 
\end{equation}
where $(i_n,\dots,i_1) \in \{0,1\}^n$, that is, we have a $+\frac{1}{2}$ spinon for $i=0$, while a $-\frac{1}{2}$ spinon for $i=1$. Let $j_{l_k}$ which satisfies $j_{l_{k+1}} - j_{l_k} = H(p_{l_{k+1}},p_{l_k})$ denoted by $j_{r_s}$ ($s=1,\dots,M$). Then we have another bijection: 
%the following isomorphism for a sequence $(r_{M,M-1}, \dots, r_{2,1}) := (r_M - r_{M-1}, \dots, r_2 - r_1)$: 
\begin{equation} \label{eq:Yangian}
\begin{split}
  &\Big\{ \varphi_{j_n}^{*p_n} \cdots \varphi_{j_1}^{*p_1} \in 
 \bigsqcup_{n=0}^{\infty} \bigsqcup_{j_1-p_1,\dots,j_n-p_n} 
 [[2j_{l_n}+i_n-p_n, \dots, 2j_{l_1}+i_n-p_1]]_{0,0}\, \Big|\, j_{l_{k+1}} - j_{l_k} = H(p_{l_{k+1}},p_{l_k}),\, j_{l_k} = j_{r_s} \Big\} \\ 
 &= B^{(r_M-r_{M-1})} \otimes \cdots \otimes B^{(r_2-r_1)}. 
\end{split}
\end{equation}
This bijection leads to the Yangian-like structure and thus we obtain the relation to Haldane's spinon orbital interpretation of spinon motifs~\cite{bib:HHTBP92}. This notion was introduced for the Haldane-Shastry (HS) spin chain~\cite{bib:H88, bib:S88}, in which the eigenstates are characterized by integer-valued rapidities. A spinon motif, which consists of parentheses and $1$'s, is defined by assigning a motif $)($ to an unoccupied integer by rapidities and $1$ to an occupied integer. If one finds a motif $(1\,\dots\,1)$ consisting of $n$ symbols by counting $)$ or $($ as a half-symbol, it represents the singlet state of $n$ spins belonging to the fundamental representations. In the case of $n=1$, a motif $()$ simply represents a single spinon state. Therefore, a sequential motif $()(\cdots)()$ consisting of $n$ symbols represents a $n$-fold tensor product of single spinon states. The irreducible decomposition of a spinon motif leads to the notion of spinon orbitals. We replace a parenthesis between symmetrized spinons by $0$. For instance, $n$ symmetrized
spinons constitute a space represented by $(0\,\dots\,0)$ with $n$ symbols. Then a distinct spinon orbital is assigned to each set of symmetrized spinons. 
This leads to the bijection between the right-hand side of \eqref{eq:Yangian} and a spinon motif as 
\begin{equation}
 B^{(r_M-r_{M-1})} \otimes \cdots \otimes B^{(r_2-r_1)} = 
  \underbrace{(0\dots0)}_{r_M-r_{M-1}} \cdots \underbrace{(0\dots0)}_{r_2-r_1}. 
\end{equation}

\section{Supersymmetry behind the ZF model} \label{sec:SUSY}
The supersymmetry of the finite ZF chain was introduced in \cite{bib:H13}. The superalgebra was defined in such a way that the supercharges change the number of system size by one. Here we modify their definitions in order to be compatible with the infinite spin chain: 
\begin{equation} \label{eq:supercharge}
 \mathsf{Q} := \sum_{j \in \mathbb{Z}} (-1)^{j-1} \mathsf{q}_j,\quad 
  \mathsf{Q}^{\dag} := \sum_{j \in \mathbb{Z}} (-1)^{j-1} \mathsf{q}_j^{\dag}, 
\end{equation}
where $\mathsf{q}_j$ and $\mathsf{q}^{\dag}_j$ nontrivially act on the $j$th site: 
\begin{equation}
\begin{split}
 &\mathsf{q}_j := 1 \otimes \dots \otimes 1 \otimes \underbrace{\mathsf{q}}_j \otimes 1 \otimes \dots \otimes 1, \\
 &\mathsf{q}^{\dag}_j := 1 \otimes \dots \otimes 1 \otimes \underbrace{\mathsf{q}^{\dag}}_j \otimes 1 \otimes \dots \otimes 1. 
\end{split}
\end{equation}
the local action $\mathsf{q}:V^{(2)} \mapsto V^{(2)} \otimes V^{(2)}$ is given by 
\begin{equation} \label{eq:local_supercharge}
\begin{split}
 &\mathsf{q}|2\rangle_j = \dfrac{q+q^{-1}}{2} (|12\rangle_{j,j+1} - |21\rangle_{j,j+1}), \\
 %\mathsf{q}_j^{\dag} \dfrac{q+q^{-1}}{2} (|12\rangle_{j,j+1} - |21\rangle_{j,j+1}) = |2\rangle_j, \\
 &\mathsf{q}|1\rangle_j = |02\rangle_{j,j+1} - |20\rangle_{j,j+1}, \\
 %\mathsf{q}_j^{\dag} (|02\rangle_{j,j+1} - |20\rangle_{j,j+1}) = |1\rangle_j, \\
 &\mathsf{q}|0\rangle_j = \dfrac{q+q^{-1}}{2} (|01\rangle_{j,j+1} - |10\rangle_{j,j+1}) 
 %\mathsf{q}_j^{\dag} \dfrac{q+q^{-1}}{2} (|01\rangle_{j,j+1} - |10\rangle_{j,j+1}) = |0\rangle_j. 
\end{split}
\end{equation}
and $\mathsf{q}^{\dag}_j:V^{(2)} \otimes V^{(2)} \mapsto V^{(2)}$ is the adjoint operator of $\mathsf{q}$. 
%The other states are killed by the action of $\mathsf{q}_j^{\dag}$. 
These supercharges obey a discrete analog of the superalgebra~\cite{bib:W82, bib:YF04}: 
\begin{equation} \label{eq:superalg}
 \mathsf{Q}^2 = (\mathsf{Q}^{\dag})^2 = 0, \quad
  [\mathsf{F},\,\mathsf{Q}] = \mathsf{Q}, \quad [\mathsf{F},\,\mathsf{Q}^{\dag}] = -\mathsf{Q}^{\dag}, 
\end{equation}
where $\mathsf{F}$ is the fermion number operator. 

The infinite ZF chain is a supersymmetric model in the sense of \cite{bib:H13}, since the Hamiltonian density in \eqref{eq:H_spin1} meets the criteria of the supersymmetry: 
\begin{equation} \label{eq:local_H}
 h_{j,j+1} = -(\mathsf{q}_j^{\dag} \otimes 1)(1 \otimes \mathsf{q}_j) - (1 \otimes \mathsf{q}_j^{\dag})(\mathsf{q}_j \otimes 1) 
  + \mathsf{q}_j \mathsf{q}_j^{\dag} + \frac{1}{2} (\mathsf{q}_j^{\dag} \mathsf{q}_j \otimes 1 + 1 \otimes \mathsf{q}_j^{\dag} \mathsf{q}_j). 
\end{equation}
This form of the local Hamiltonian guarantees that the whole Hamiltonian \eqref{eq:H_spin1} is obtained as the anticommutator of the supercharges $\{\mathsf{Q},\,\mathsf{Q}^{\dag}\} = H$.

\subsection{Supercharges acting on the spinon basis}
Although the supercharges \eqref{eq:supercharge} insert/remove a site to/from the system, they do not change the total magnetization. This motivates us to investigate how they act on the spinon basis since, with the conserved total magnetization, the number of spinons is either conserved or pairs of $+\frac{1}{2}$ spinons and $-\frac{1}{2}$ spinons are generated. 
We normalize the local supercharges as $\tilde{\mathsf{q}}_j = q \mathsf{q}_j$ in order not to diverge at $q \to 0$. Then the actions given in \eqref{eq:local_supercharge} are modified into 
\begin{equation} \label{eq:local_supercharge0}
\begin{split}
 &\tilde{\mathsf{q}}|2\rangle_j = \dfrac{1}{2} (|12\rangle_{j,j+1} - |21\rangle_{j,j+1}), \\
 %\dfrac{\tilde{\mathsf{q}}_j^{\dag}}{2} (|12\rangle_{j,j+1} - |21\rangle_{j,j+1}) = |2\rangle_j, \\
 &\tilde{\mathsf{q}}|1\rangle_j = 0, \\
 &\tilde{\mathsf{q}}|0\rangle_j = \dfrac{1}{2} (|01\rangle_{j,j+1} - |10\rangle_{j,j+1}). 
 %\dfrac{\tilde{\mathsf{q}}_j^{\dag}}{2} (|01\rangle_{j,j+1} - |10\rangle_{j,j+1}) = |0\rangle_j. 
\end{split}
\end{equation}

It is obtained from the direct calculation that the supercharge $\tilde{\mathsf{Q}}$ kills the $(0,0)$ ground state paths: 
\begin{equation}
\begin{split}
 \tilde{\mathsf{Q}}[[\,]]_{0,0} &= \sum_{j \in \mathbb{Z}} (-1)^{j-1} \mathsf{q}_j |\dots 0202 \dots\rangle \\
 &= \dots - \frac{1}{2} \left( |\dots 01202 \dots\rangle - |\dots 10202 \dots\rangle \right)
 + \frac{1}{2} \left( |\dots 01202 \dots\rangle - |\dots 02102 \dots\rangle \right) \\
 &\hspace{9mm}- \frac{1}{2} \left( |\dots 02012 \dots\rangle - |\dots 02102 \dots\rangle \right)
 + \frac{1}{2} \left( |\dots 02012 \dots\rangle - |\dots 02021 \dots\rangle \right) - \dots \\
 &= 0. 
\end{split}
\end{equation}
Similarly, the $(2,2)$ ground state path is killed by $\tilde{\mathsf{Q}}: [[\,]]_{2,2} \mapsto 0$. The $(1,1)$ ground state path vanishes from the definition of the local action of $\mathsf{q}$ on $|1\rangle$. 

A path with a single domain wall is given by the $(m,m')$ ground state path with $|m-m'| = 1$. 
%We denote an $(m,m \pm 1)$ ground state paths by $[[l_1]]_{m,m \pm 1}$. 
The supercharge $\tilde{\mathsf{Q}}$ defined through the local relation \eqref{eq:local_supercharge0} acts on a path with a single domain wall as 
\begin{equation} \label{eq:Q_action}
\begin{split}
 &\tilde{\mathsf{Q}} [[l_1]]_{0,1} = -\dfrac{1}{2} [[l_1+1]]_{2,1}, \quad
 \tilde{\mathsf{Q}} [[l_1]]_{2,1} = \dfrac{1}{2} [[l_1+1]]_{0,1}, \\
 &\tilde{\mathsf{Q}} [[l_1]]_{1,0} = \dfrac{1}{2} [[l_1]]_{1,0}, \quad
 \tilde{\mathsf{Q}} [[l_1]]_{1,2} = -\dfrac{1}{2} [[l_1]]_{1,2}. 
\end{split}
\end{equation}
%\begin{equation} 
%\begin{split}
% &\tilde{\mathsf{Q}} [[j]]_{1,2} = \tilde{\mathsf{Q}} \varphi_{j}^{*1}[[\,]] = -\dfrac{1}{2} \varphi_{j}^{*1}[[\,]] = -\dfrac{1}{2} [[j]]_{1,2}, \\
% &\tilde{\mathsf{Q}} [[j]]_{1,0} = \tilde{\mathsf{Q}} \varphi_{j}^{*0}[[\,]] = \dfrac{1}{2} \varphi_{j}^{*0}[[\,]] = \dfrac{1}{2} [[j]]_{1,0}, \\
% &\tilde{\mathsf{Q}} [[j]]_{2,1} = \tilde{\mathsf{Q}} \varphi_{j}^{*0}[[\,]] = \dfrac{1}{2} \varphi_{j+2}^{*1}[[\,]] = \dfrac{1}{2} [[j+1]]_{0,1}, \\
% &\tilde{\mathsf{Q}} [[j]]_{0,1} = \tilde{\mathsf{Q}} \varphi_{j}^{*1}[[\,]] = -\dfrac{1}{2} \varphi_{j}^{*0}[[\,]] = -\dfrac{1}{2} [[j+1]]_{2,1}. 
%\end{split}
%\end{equation}
These are interpreted as the supercharge locally acts on the $r$th operator of a sequence of $n$ spinon creation operators \eqref{eq:spinon_state} as 
\begin{equation}
\begin{split}
 &\tilde{\mathsf{Q}} (\varphi_{j_r}^{*1}) = -\dfrac{1}{2} \varphi_{j_r}^{*}, \quad
 ((-1)^{p_1} + \cdots + (-1)^{p_{r-1}} = 1) \land (p_r = 1) \\
 &\tilde{\mathsf{Q}} (\varphi_{j_r}^{*0}) = \dfrac{1}{2} \varphi_{j_r+2}^{*1}, \quad
 ((-1)^{p_1} + \cdots + (-1)^{p_{r-1}} = 1) \land (p_r = 0) \\
 &\tilde{\mathsf{Q}} (\varphi_{j_r}^{*0}) = \dfrac{1}{2} \varphi_{j_r}^{*0}, \quad
 ((-1)^{p_1} + \cdots + (-1)^{p_{r-1}} = 0) \land (p_r = 0) \\
 &\tilde{\mathsf{Q}} (\varphi_{j_r}^{*1}) = -\dfrac{1}{2} \varphi_{j_r}^{*1}, \quad
 ((-1)^{p_1} + \cdots + (-1)^{p_{r-1}} = 2) \land (p_r = 1). 
\end{split}
\end{equation}

In order to define the sequential action of the supercharge on $n$ spinon creation operators, we introduce the operator $\mathsf{\Gamma}$. This operator acts on a path with a single domain wall as 
\begin{equation} \label{eq:Gamma_action}
\begin{split}
 &\mathsf{\Gamma} [[l_1]]_{0,1} = [[l_1+1]]_{2,1}, \quad
 \mathsf{\Gamma} [[l_1]]_{2,1} = [[l_1+1]]_{0,1}, \\
 &\mathsf{\Gamma} [[l_1]]_{1,0} = [[l_1-1]]_{1,2}, \quad
 \mathsf{\Gamma} [[l_1]]_{1,2} = [[l_1-1]]_{1,0}, 
\end{split}
\end{equation}
which leads to the local action on the $r$th operator of a sequence of $n$ spinon operators as 
\begin{equation} 
\begin{split}
 &\mathsf{\Gamma} (\varphi_{j_r}^{*1}) = \varphi_{j_r}^{*0}, \quad
 ((-1)^{p_1} + \cdots + (-1)^{p_{r-1}} = 1) \land (p_r = 1) \\
 &\mathsf{\Gamma} (\varphi_{j_r}^{*0}) = \varphi_{j_r+2}^{*1}, \quad
 ((-1)^{p_1} + \cdots + (-1)^{p_{r-1}} = 1) \land (p_r = 0) \\
 &\mathsf{\Gamma} (\varphi_{j_r}^{*0}) = \varphi_{j_r}^{*1}, \quad
 ((-1)^{p_1} + \cdots + (-1)^{p_{r-1}} = 0) \land (p_r = 0) \\
 &\mathsf{\Gamma} (\varphi_{j_r}^{*1}) = \varphi_{j_r-2}^{*0}, \quad
 ((-1)^{p_1} + \cdots + (-1)^{p_{r-1}} = 2) \land (p_r = 1). 
\end{split}
\end{equation}
%Similarly in \eqref{eq:Q_action}, the above $\mathsf{\Gamma}$ actions are interpreted as 
%\begin{equation} \label{eq:Gamma_action}
% \mathsf{\Gamma}(\varphi_{j}^{*1}) = \varphi_{j-2}^{*0}, \quad
%  \mathsf{\Gamma}(\varphi_{j}^{*0}) = \varphi_{j}^{*1}, \quad
%  \mathsf{\Gamma}(\varphi_{j}^{*0}) = \varphi_{j+2}^{*1}, \quad
%  \mathsf{\Gamma}(\varphi_{j}^{*1}) = \varphi_{j}^{*0}. 
%\end{equation}
Note that $\tilde{\mathsf{Q}}$ and $\mathsf{\Gamma}$ do not change the spinon spins since the subscripts are invariant under their actions in mod $2$. The number of spinons is also conserved and therefore, they only act on the domain-labels, denoted by the superscripts. 

We obtain that the actions of the supercharges $\tilde{\mathsf{Q}}$, $\tilde{\mathsf{Q}}^{\dag}$ on a sequence of $n$ spinon creation operators is written by 
\begin{equation} \label{eq:Qaction_multi}
\begin{split}
 &\tilde{\mathsf{Q}} (\varphi_{j_n}^{*p_n} \dots \varphi_{j_n}^{*p_n}) 
 = \tilde{\mathsf{Q}} (\varphi_{j_n}^{*p_n}) 1 (\varphi_{j_{n-1}}^{*p_{n-1}}) \cdots 1 (\varphi_{j_1}^{*p_1}) 
 + \mathsf{\Gamma} (\varphi_{j_n}^{*p_n}) \tilde{\mathsf{Q}} (\varphi_{j_{n-1}}^{*p_{n-1}}) 1 (\varphi_{j_{n-2}}^{*p_{n-2}}) \cdots 1 (\varphi_{j_1}^{*p_1}) \\
 &\hspace{27mm}
 + \cdots + \mathsf{\Gamma} (\varphi_{j_n}^{*p_n}) \dots \mathsf{\Gamma} (\varphi_{j_2}^{*p_2}) \cdots \tilde{\mathsf{Q}} (\varphi_{j_1}^{*p_1}), \\
 &\tilde{\mathsf{Q}}^{\dag} (\varphi_{j_n}^{*p_n} \dots \varphi_{j_n}^{*p_n}) 
 = \tilde{\mathsf{Q}}^{\dag} (\varphi_{j_n}^{*p_n}) 1 (\varphi_{j_{n-1}}^{*p_{n-1}}) \cdots 1 (\varphi_{j_1}^{*p_1}) 
 + \mathsf{\Gamma} (\varphi_{j_n}^{*p_n}) \tilde{\mathsf{Q}}^{\dag} (\varphi_{j_{n-1}}^{*p_{n-1}}) 1 (\varphi_{j_{n-2}}^{*p_{n-2}}) \cdots 1 (\varphi_{j_1}^{*p_1}) \\
 &\hspace{27mm}
 + \cdots + \mathsf{\Gamma} (\varphi_{j_n}^{*p_n}) \dots \mathsf{\Gamma} (\varphi_{j_2}^{*p_2}) \cdots \tilde{\mathsf{Q}}^{\dag} (\varphi_{j_1}^{*p_1}). 
\end{split}
\end{equation}
The actions \eqref{eq:Q_action} and \eqref{eq:Gamma_action} are expressed in terms of four-by-four matrices, if we choose the basis $(m_r,m_{r-1})$ as $\{(0,1),(2,1),(1,0),(1,2)\}$: 
\begin{equation} \label{eq:supergen_matrix}
 \tilde{\mathsf{Q}} = \frac{1}{2}
  \begin{pmatrix}
   0 & 1 & 0 & 0 \\ -1 & 0 & 0 & 0 \\ 0 & 0 & 1 & 0 \\ 0 & 0 & 0 & -1
  \end{pmatrix}, \quad
  \tilde{\mathsf{Q}}^{\dag} = \frac{1}{2}
  \begin{pmatrix}
   0 & -1 & 0 & 0 \\ 1 & 0 & 0 & 0 \\ 0 & 0 & 1 & 0 \\ 0 & 0 & 0 & -1
  \end{pmatrix}, \quad
  \mathsf{\Gamma} = 
  \begin{pmatrix}
   0 & 1 & 0 & 0 \\ 1 & 0 & 0 & 0 \\ 0 & 0 & 0 & 1 \\ 0 & 0 & 1 & 0
  \end{pmatrix}. 
\end{equation}
These are related to the generators $Q$, $\bar{Q}$, $\Gamma$ of the superalgebra introduced for the supersymmetric sine-Gordon model~\cite{bib:HM97} via 
\begin{equation} 
 \tilde{\mathsf{Q}} = \dfrac{e^{\pi i/4}}{2\sqrt{2}} (Q - i\bar{Q}), \quad
  \tilde{\mathsf{Q}}^{\dag} = \dfrac{e^{-\pi i/4}}{2\sqrt{2}} (Q + i\bar{Q}), \quad
  \mathsf{\Gamma} = \Gamma. 
\end{equation}
Moreover, the action of $\tilde{\mathsf{Q}}$, $\tilde{\mathsf{Q}}^{\dag}$ on the sequential operators \eqref{eq:Qaction_multi} coincides with the comultiplication of the supercharge~\cite{bib:S90}: 
\begin{equation} \label{eq:comultiplication}
\begin{split}
 &\Delta^{(n)}(Q) = \sum_{j=1}^n
 %\underbrace{Q}_n \otimes 1 \otimes \dots \otimes 1 
 \Gamma \otimes \dots \otimes \Gamma \otimes \underbrace{Q}_j \otimes 1 \otimes \dots \otimes 1, \\
 %+ \dots + \Gamma \otimes \dots \otimes \Gamma \otimes \underbrace{Q}_2 \otimes 1 
 %+ \Gamma \otimes \dots \otimes \Gamma \otimes \underbrace{Q}_1, \\
 &\Delta^{(n)}(\bar{Q}) = \sum_{j=1}^n
 %\underbrace{\bar{Q}}_n \otimes 1 \otimes \dots \otimes 1 
 \Gamma \otimes \dots \otimes \Gamma \otimes \underbrace{\bar{Q}}_j \otimes 1 \otimes \dots \otimes 1. 
 %+ \dots + \Gamma \otimes \dots \otimes \Gamma \otimes \underbrace{\bar{Q}}_2 \otimes 1
 %+ \Gamma \otimes \dots \otimes \Gamma \otimes \underbrace{\bar{Q}}_1. 
\end{split}
\end{equation}

Besides the number of spinons and each of spinon spins, the surpercharges $\tilde{\mathsf{Q}}$, $\tilde{\mathsf{Q}}^{\dag}$ do not change the sequence $(r_M-r_{M-1}, \dots, r_2-r_1)$. Therefore, the associated Haldane's spinon motifs are invariant under the actions of supercharges. 
Thus, the supercharges does not change the spinon spins {\it i.e.} the $SU(2)$-multiplet structure of spinons but nontrivially act on the RSOS part by changing the labels of domains.

\subsection{Superalgebra in spinon basis}
In the previous subsection, we have obtained the actions of the supercharges on a sequence of spinon creation operators. In this subsection, we discuss which subspace of the Hilbert space of the FZ chain is supersymmetric. In \cite{bib:H13}, the discrete analog of the supersymmetry was discussed only in the subspace with limit momentum, {\it i.e.} the zero momentum space for odd $N$ and the $\pi$ momentum space for even $N$. In this subspace, we have no restriction from $E \geq |P|$. 
In order to determine the superspace for which the supersymmetry holds, we study the superalgebra \eqref{eq:superalg} in the spinon basis.  
%give the subspace spanned by the basis for which the superalgebra \eqref{eq:superalg} is satisfied by the matrices \eqref{eq:supergen_matrix}. At the same time, we discuss which basis forms the eigenspace of the Hamiltonian defined through the supercharges \eqref{eq:local_H}. 

%The verification of the matrix expressions \eqref{eq:supergen_matrix} is achieved by checking the superalgebra \eqref{eq:superalg}. 
Taking into account that $(-1)^{\mathsf{F}} = \mathsf{\Gamma}$, the second and third relations of \eqref{eq:superalg} are rewritten as 
\begin{equation} \label{eq:superalg2}
 \{\mathsf{\Gamma},\, \mathsf{\tilde{Q}}\} = \{\mathsf{\Gamma},\, \mathsf{\tilde{Q}}^{\dag}\} = 0. 
\end{equation}
These relations are checked through direct calculation in the matrix forms \eqref{eq:supergen_matrix}. The first relation of \eqref{eq:superalg} is proved through the coproductive forms \eqref{eq:Qaction_multi}. Noting that 
\begin{equation} \label{eq:squares}
 \tilde{\mathsf{Q}}^2 = (\tilde{\mathsf{Q}}^{\dag})^2 = \frac{1}{4}
  \begin{pmatrix}
   -1 & 0 & 0 & 0 \\ 0 & -1 & 0 & 0 \\ 0 & 0 & 1 & 0 \\ 0 & 0 & 0 & 1
  \end{pmatrix}, \quad
  \mathsf{\Gamma}^2 = 1 
\end{equation}
and the anticommutation relations \eqref{eq:superalg2}, the squares of the coproductive supercharges act on a sequence of $n$ spinon creation operators as 
\begin{equation} \label{eq:reduced_squares}
\begin{split}
 &(\Delta^{(n)}(\tilde{\mathsf{Q}}))^2 = \sum_{j=1}^n 1 \otimes \cdots \otimes 1 \otimes \underbrace{\tilde{\mathsf{Q}}^2}_j \otimes 1 \cdots \otimes 1, \\
 &(\Delta^{(n)}(\tilde{\mathsf{Q}^{\dag}}))^2 = \sum_{j=1}^n 1 \otimes \cdots \otimes 1 \otimes \underbrace{(\tilde{\mathsf{Q}}^{\dag})^2}_j \otimes 1 \cdots \otimes 1. 
\end{split}
\end{equation}
The summations vanish only for even $n$. Thus, the supersymmetry is obtained only when an even number of spinons exists, that is, the infinite FZ chain possesses supersymmetric subspace, which is spanned by paths with boundary conditions $m_0 = 0$, $m_n = 0,2$.

\subsection{Graphical representations of restricted paths}
\begin{figure}
 \centering
 \includegraphics[scale=0.5]{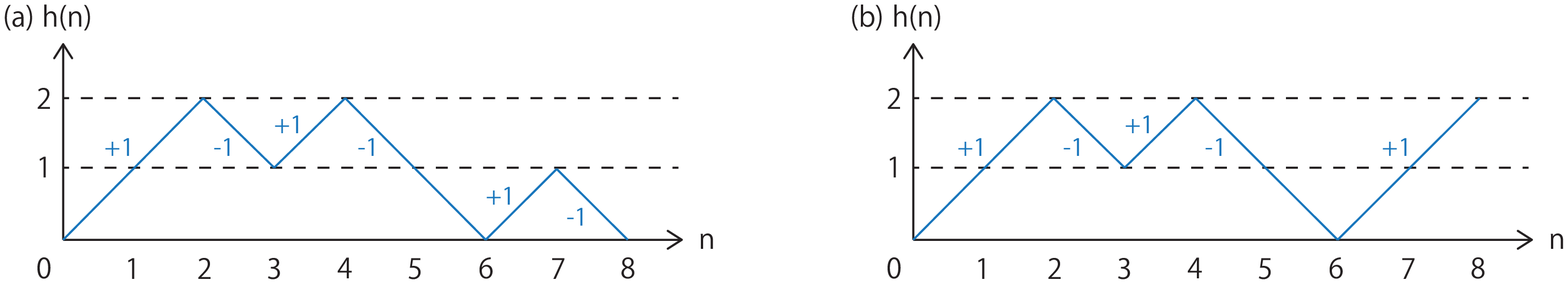}
 \caption{Two examples of the restricted path with $n = 8$. (a) The restricted path from $0$ to $0$ under the boundary conditions $m_0=0,\,m_8=0$. This path is given by the set $(p_8,p_7,p_6,p_5,p_4,p_3,p_2,p_1) = (1,0,1,1,0,1,0,0)$. The local maxima and minima characterizing the restricted path are obtained as $(r_2,r_1) = (7,3)$ and $(r_5^{\rm tot},r_4^{\rm tot},r_3^{\rm tot},r_2^{\rm tot},r_1^{\rm tot}) = (7,6,4,3,2)$. (b) The restricted path from $0$ to $2$ under the boundary conditions $m_0=0,\,,m_8 = 2$. This path is given by the set $(p_8,p_7,p_6,p_5,p_4,p_3,p_2,p_1) = (0,0,1,1,0,1,0,0)$. The local maxima and minima characterizing the restricted path are obtained as $r_1 = 3$ and $(r_4^{\rm tot},r_3^{\rm tot},r_2^{\rm tot},r_1^{\rm tot}) = (6,4,3,2)$.} \label{fig:res_path_def}
\end{figure}

Here we give the graphical representations of a restricted path, which represents the RSOS degrees of freedom of the FZ chain. A restricted path is encoded by a sequence $(p_n,\dots,p_1)$ ($p_j = 0,1$) satisfying the restriction \eqref{eq:res_path}. For a fixed sequence of $p_j$'s, set 
\begin{equation}
 h(n) = (-1)^{p_1} + \cdots + (-1)^{p_n}
\end{equation}
as a function of $n$. Then we have a graphical representation of the restricted path by taking $n$ on the $x$-axis and $h(n)$ on the $y$-axis (Fig.~\ref{fig:res_path_def}). Thus, $h(n)$ is considered as a height function. The adjacency condition \eqref{eq:ad_cond2} indicates that a graph consists only of $\pm 1$ slopes. Since the paths which span the supersymmetric subspace of the FZ model meet the boundary conditions $m_0 = 0$, $m_n = 0,2$, the associated restricted paths start from $(0,0)$ and end with $(n,0)$ or $(n,2)$ on a graph. Note that, in a $(m_n,m_0)$ path, the value of the function $h(j)$ reads the domain after the $j$th domain wall. 

\begin{figure}
 \centering
 \includegraphics[scale=0.5]{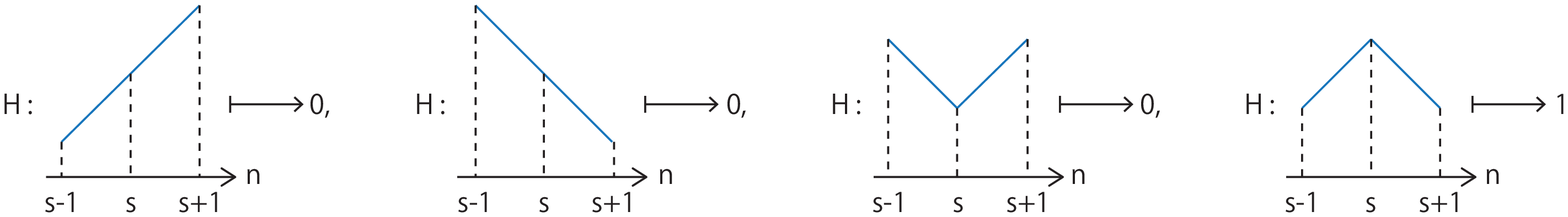}
 \caption{The values of the energy function applied to restricted paths of length two.} \label{fig:energy_func}
\end{figure}

Here we remark properties of a restricted path. 
%The graphical representation makes the discussion about these properties easier. 
Let us denote $j$ which satisfies 
\begin{equation} \label{eq:height1}
 h(j) = 1 \land
  h(j-1) = h(j+1)
\end{equation}
by $r_s$. This is interpreted as the existence of local maxima/minima with height $1$ in a graph. Due to the condition \eqref{eq:res_path}, a restricted path of length $n$ is uniquely determined by a set of $r_s$'s under a fixed boundary conditions. 
We write a restricted path of length $n$ characterized by $(p_{r_M}, \dots, p_{r_1})$ $(r_1 < \dots < r_M)$ as $p^{\rm res}_{n}(p_{r_M}, \dots, p_{r_1})$. Due to the boundary conditions, any $r_s$ takes an odd-integer value. The adjacency condition imposes that we have $M \leq \frac{n}{2}$, which is an integer since the supersymmetry is obtained only for even $n$. Once $M$ is fixed, the number of $j$'s which satisfy $h(j-1) = h(j+1)$ is uniquely determined: 
\begin{equation}
\begin{split}
 \# r^{\rm tot}_s &:= 
 \# \{j|h(j-1) = h(j+1)\}\\ &= 
  \begin{cases}
   \frac{1}{2}(n - 2M) + 2M - 1 & m_n = 0 \\
   \frac{1}{2}(n - 2M - 2) + 2M & m_n = 2. 
  \end{cases}
\end{split}
\end{equation}
Note that $r^{\rm tot}_s$'s read the positions of local maximum/minimum of a graph. 
%The corresponding object to $N^{\rm tot}$ in the path is the number of fragments of groundstate paths after removing blank paths. 
Taking all these into consideration, we obtain that there are $\begin{pmatrix} \frac{n}{2} \\ M  \end{pmatrix}$ distinct paths. 

\begin{figure}
 \centering
 \includegraphics[scale=0.5]{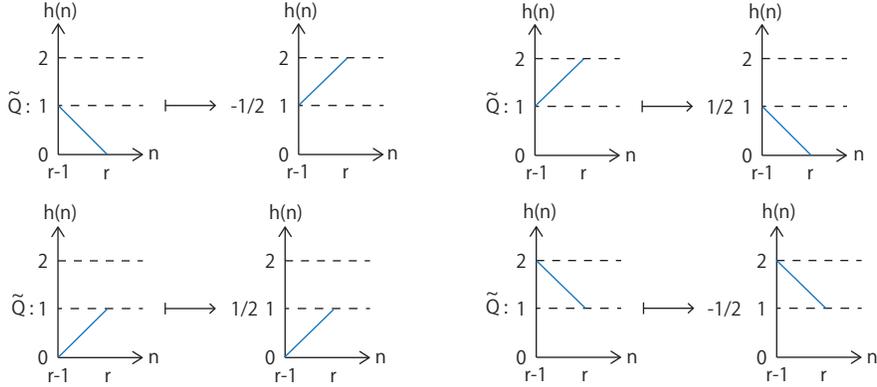}
 \caption{The actions of the supercharges on each restricted path.} \label{fig:supercharge}
\end{figure}

\begin{figure}
 \centering
 \includegraphics[scale=0.35]{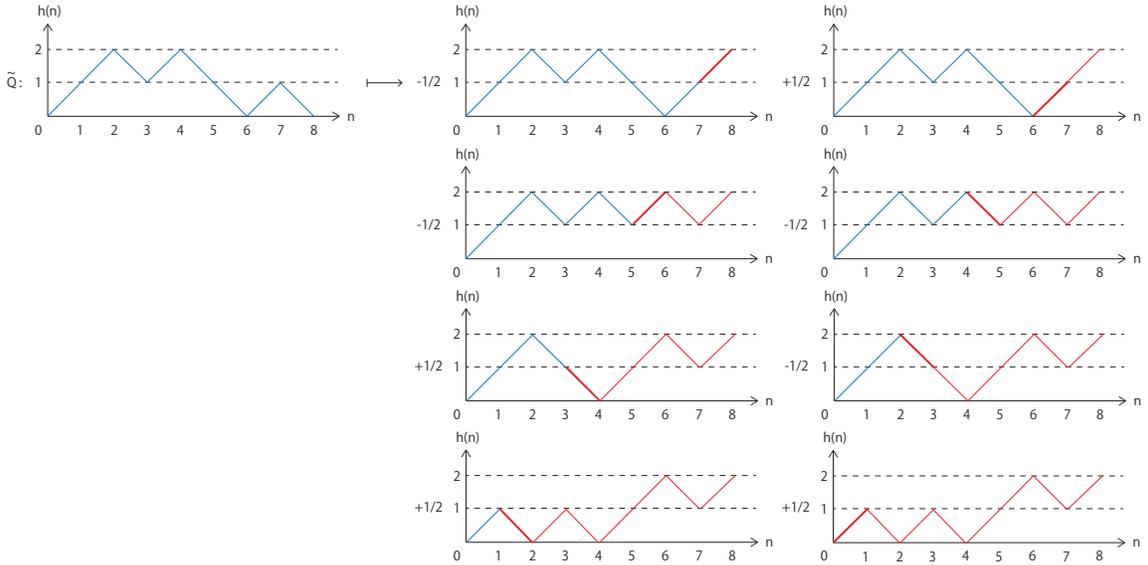}
 \caption{The coproductive action of the supercharge on the restricted path with $n = 8$. The bold red lines represent where $\tilde{\mathsf{Q}}$ locally acts and the thin red lines show where $\mathsf{\Gamma}$ acts. Each two terms in the first and third lines compensate, resulting in zero.} \label{fig:comultiplication}
\end{figure}

The energy function $H(p_s,p_{s-1})$ in \eqref{eq:energy_func} is graphically represented as in Fig.~\ref{fig:energy_func}. The function maps a path of length two either $0$ or $1$. The supercharge $\tilde{\mathsf{Q}}$ also admits graphical actions (Fig.~\ref{fig:supercharge}). These are determined from the matrix expressions \eqref{eq:supergen_matrix}. Similarly, the graphical actions of $\tilde{\mathsf{Q}}^{\dag}$ and $\mathsf{\Gamma}$ are determined. 
%Since $h(j)$ indicates the local groundstate path after the $j$th domain wall, the graphical representation is directly connected to the path basis in \eqref{eq:supergen_matrix}. 
One obtains that the operator $\mathsf{\Gamma}$ simply exchanges the labels $0$ and $2$. The comultiplication \eqref{eq:comultiplication} is realized subsequently by applying these local actions to the restricted path. We show the $n=8$ examples in Fig.~\ref{fig:comultiplication}. 

\subsection{Construction of superspace}
In the supersymmetric model, every eigenstate except for the ground state makes a superpartner by sharing the same energy. A superpartner consists of the eigenstates which are mapped to each other by $\tilde{\mathsf{Q}}$ and $\tilde{\mathsf{Q}}^{\dag}$. Due to the nilpotency of the supercharges, these states are annihilated either by $\tilde{\mathsf{Q}}$ or by $\tilde{\mathsf{Q}}^{\dag}$. This property is a straightforward consequence of the definition of the Hamiltonian as the anticommutator $\{\tilde{\mathsf{Q}},\,\tilde{\mathsf{Q}}^{\dag}\}$. 
Although the Hamiltonian with the Hamiltonian density given by \eqref{eq:local_H} coincides with the known FZ Hamiltonian \eqref{eq:H_spin1}, \eqref{eq:coupling_const}, any of paths are invariant under the $q \to 0$ limit of the latter Hamiltonian but not invariant under the former one. This is understood as due to degeneracy occurring in the $q \to 0$ limit. We expect that the structure of linearly combined paths invariant under the supersymmetry holds for arbitrary anisotropy, which is one of our future problems. 
%This will help us to study the spin chains with spin-$\frac{\ell}{2}$, which have the supersymmetry only at the combinatorial point $q = e^{\frac{\pi i}{\ell+2}}$. Besides, the structure may hold for the finite size system, since the Yangian-like structure seems be obtained for the finite spin-$\frac{1}{2}$ chain~\cite{bib:AKMW06}. 
After some manipulation, we found that a certain linear combination of paths constitutes the invariant subspace of both of the above Hamiltonians. 

%Therefore, the aim of this subsection is to find a proper linear combination of paths which is invariant under the anticommutator of the supercharges. Note that, all eigenstates of the supersymmetric model make superpartners except for the ground state. 
Since any of invariant paths are killed by one of two terms constituting the anticommutator $\{\tilde{\mathsf{Q}},\,\tilde{\mathsf{Q}}^{\dag}\}$, we look up these two actions on a path separately. 
Let us first consider the action of the product $\tilde{\mathsf{Q}} \tilde{\mathsf{Q}}^{\dag}$. By expanding them in the coproductive forms, we have 
\begin{equation} \label{eq:exp_comulti}
\begin{split}
 \Delta^{(n)}(\tilde{\mathsf{Q}}) \Delta^{(n)}(\tilde{\mathsf{Q}}^{\dag}) &= 
 \sum_{j > i} 1 \otimes \cdots \otimes 1 \otimes \underbrace{\mathsf{\Gamma} \tilde{\mathsf{Q}}^{\dag}}_j \otimes \mathsf{\Gamma} \otimes \cdots \otimes \mathsf{\Gamma} \otimes \underbrace{\tilde{\mathsf{Q}}}_i \otimes 1 \otimes \cdots \otimes 1 \\ 
 &+ \sum_{j > i} 1 \otimes \cdots \otimes 1 \otimes \underbrace{\tilde{\mathsf{Q}} \mathsf{\Gamma}}_j \otimes \mathsf{\Gamma} \otimes \cdots \otimes \mathsf{\Gamma} \otimes \underbrace{\tilde{\mathsf{Q}}^{\dag}}_i \otimes 1 \otimes \cdots \otimes 1 \\
 &+ \sum_{i=1}^n 1 \otimes \cdots \otimes 1. 
\end{split} 
\end{equation}
%the first and second terms. $\mathsf{\Gamma}$ simply exchanges the domains $0$ and $2$
The actions of $\tilde{\mathsf{Q}}$, $\tilde{\mathsf{Q}}^{\dag}$, and $\mathsf{\Gamma}$ on a restricted path are given in \eqref{eq:supergen_matrix}, while we have 
\begin{equation}
 \mathsf{\Gamma} \tilde{\mathsf{Q}}^{\dag} = \frac{1}{2}
  \begin{pmatrix} 
   1 & 0 & 0 & 0 \\ 0 & -1 & 0 & 0 \\ 0 & 0 & 0 & -1 \\ 0 & 0 & 1 & 0
  \end{pmatrix}, \quad
  \tilde{\mathsf{Q}} \mathsf{\Gamma} = \frac{1}{2}
  \begin{pmatrix}
   1 & 0 & 0 & 0 \\ 0 & -1 & 0 & 0 \\ 0 & 0 & 0 & 1 \\ 0 & 0 & -1 & 0
  \end{pmatrix}. 
\end{equation} 
Thus, the actions of the first and second terms in \eqref{eq:exp_comulti} cancel each other out unless $j-i \in 2\mathbb{Z}_{\geq 0}+1$. 

For a sequence $(\check{r}_1,\dots,\check{r}_{\frac{n}{2}-M-1})$ ($\check{r}_j \in \{1,3,\dots,n-1\} \backslash \{r_1,\dots,r_M\}$), we define a set of $s_t$'s by $\mathcal{S}_j := \bigsqcup_{t;r_{j-1} < s_t < r_j} s_t$. If there exists no such $s_t$, then $\mathcal{S}_j = \emptyset$. 
From direct calculation, we obtain 
\begin{equation} \label{eq:QdagQ}
\begin{split}
 &\tilde{\mathsf{Q}} \tilde{\mathsf{Q}}^{\dag} p^{\rm res}_{n}(p_{r_M}, \dots, p_{r_1}) \\
 &= 
 \frac{n}{4} p^{\rm res}_{n}(p_{r_M}, \dots, p_{r_1})
 + \left\{\left(\frac{n}{2} - M\right) \left(\frac{1}{2}\right) + M \left(-\frac{1}{2}\right)\right\} p^{\rm res}_{n}(p_{r_M}, \dots, p_{r_1}) \\
 &+ \sum_{j=1}^M \sum_{j'=1}^M \sum_{s_t \in \mathcal{S}_{j'}}
 (-1)^{\frac{1}{2}|r_j-s_t| - \lfloor| j-j'+\frac{1}{2} |\rfloor}
 p_n^{\rm res}(p_{\sigma(r_{M})},\dots,\check{p}_{\sigma(r_j)},p_{\sigma(s_t)},\dots,p_{\sigma(r_1)})
% &+ \sum_{\check{p}_{r_j} \in \{p_{r_1},\dots,p_{r_M}\} \atop p_{r'_j} \in \{1,3,\dots,\check{p}_{r_j}-2\} \backslash \{p_{r_1},\dots,p_{r_j-1}\}}
% (-1)^{\frac{1}{2}(\check{p}_{r_j} - p_{r'_j}) - \frac{1}{2} \sum_{k=1}^{j-1}(\check{p}_{r_j} - p_{r_k})}
% p^{\rm res}_{n}(p_{\sigma(r_M)}, \dots, p_{\sigma(r'_j)}, \check{p}_{r_j}, \dots, p_{\sigma(r_1)}) \\
% &+ \sum_{\check{p}_{r_j} \in \{p_{r_1},\dots,p_{r_M}\} \atop p_{r'_j} \in \{\check{p}_{r_j}+2,\dots,N_s-3,N_s-1\} \backslash \{p_{r_j+1},\dots,p_{r_M}\}} 
% (-1)^{\frac{1}{2}(p_{r'_j} - \check{p}_{r_j}) - \frac{1}{2} \sum_{k=j+1}^{M}(p_{r_k} - \check{p}_{r_j})}
% p^{\rm res}_{n}(p_{\sigma(r_M)}, \dots, \check{p}_{r_j}, p_{\sigma(r'_j)}, \dots, p_{\sigma(r_1)}), 
\end{split}
\end{equation}
where $\check{p}_{r_j}$ indicates a local minimum or maximum removed and replaced by $p_{s_t}$. We used the notation $\lfloor * \rfloor$ in order to represent the largest integer no greater than $*$. The first term comes from the last term of \eqref{eq:exp_comulti}, the second term comes from the sums over odd $j$, and the rest comes from the sums over even $j$. 
Note that $p_{s_t}$ does not always satisfy $r_{j+1} > s_t > r_{j-1}$. For this reason, we introduced $\sigma$ for reordering the indices $r_M,\dots,s_t,\dots,r_1$ in such a way that $\sigma(r_M) > \dots > \sigma(s_t) > \dots > \sigma(r_1)$. 

\begin{figure}
 \centering
 \includegraphics[scale=0.55]{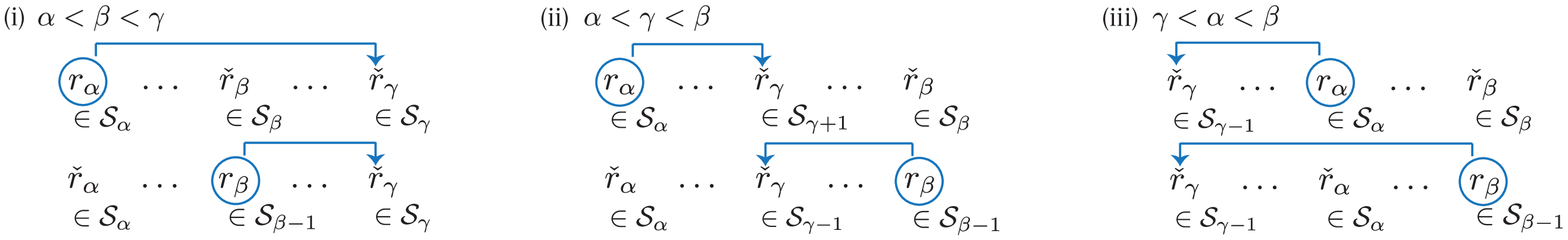}
 \caption{The possible configurations of $r_{\alpha},\,r_{\beta},\, r_{\gamma}$ and the sets they belong to. In all cases (i)-(iii), the upper one represents a map $\tilde{\mathsf{Q}} \tilde{\mathsf{Q}}^{\dag}$ replacing $r_{\alpha}$ with $\check{r}_{\gamma}$, while the lower one represents a map replacing $r_{\beta}$ with $\check{r}_{\gamma}$. 
(i) The upper one gets $(-1)^{-\frac{1}{2}(r_{\alpha}-\check{r}_{\gamma}) - (\alpha-\gamma+1) - \frac{1}{2}(\check{r}_{\beta}-r_1) + (\beta-1)}$ and the lower one gets $(-1)^{-\frac{1}{2}(r_{\beta}-\check{r}_{\gamma}) - (\beta-\gamma) - \frac{1}{2}(\check{r}_{\alpha}-r_1) + (\alpha-1)}$. 
(ii) The upper one gets $(-1)^{-\frac{1}{2}(r_{\alpha}-\check{r}_{\gamma}) - (\alpha-\gamma+1) - \frac{1}{2}(\check{r}_{\beta}-r_1) + (\beta-1)}$ and the lower one gets $(-1)^{\frac{1}{2}(r_{\beta}-\check{r}_{\gamma}) + (\beta-\gamma) - \frac{1}{2}(\check{r}_{\alpha}-r_1) + (\alpha-1)}$. 
(iii) The upper one gets $(-1)^{\frac{1}{2}(r_{\alpha}-\check{r}_{\gamma}) + (\alpha-\gamma+1) - \frac{1}{2}(\check{r}_{\beta}-r_1) + (\beta-1)}$ and the lower one gets $(-1)^{\frac{1}{2}(r_{\beta}-\check{r}_{\gamma}) + (\beta-\gamma) - \frac{1}{2}(\check{r}_{\alpha}-r_1) + (\alpha-1)}$. Thus we obtain that, in all cases, the upper one and the lower one result in the opposite signs. } \label{fig:QQdag}
\end{figure}

\begin{figure}
 \centering
 \includegraphics[scale=0.55]{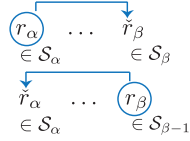}
 \caption{Under a map $\tilde{\mathsf{Q}} \tilde{\mathsf{Q}}^{\dag}$, the upper one gets $(-1)^{-\frac{1}{2}(r_{\alpha}-r_{\beta}) - (\alpha-\beta+1) - \frac{1}{2}(r_{\beta}-1) + (\beta-1)}$ and the lower one gets $(-1)^{\frac{1}{2}(r_{\beta}-r_{\alpha}) + (\beta-\alpha-1) - \frac{1}{2}(r_{\alpha}-1) + (\alpha-1))}$, which result in the same signs. } \label{fig:QQdag2}
\end{figure}

What \eqref{eq:QdagQ} indicates is as follows. Suppose we have two restricted paths 
$p_n^{\rm res}(p_{\sigma(r_{\alpha})},p_{\sigma(r_{M-1})},\dots,p_{\sigma(r_1)})$ and 
$p_n^{\rm res}(p_{\sigma(r_{\beta})},p_{\sigma(r_{M-1})},\dots,p_{\sigma(r_1)})$. 
The actions $\tilde{\mathsf{Q}} \tilde{\mathsf{Q}}^{\dag}$ on these restricted paths produce the same restricted path 
$p_n^{\rm res}(p_{\sigma(r_{\gamma})},p_{\sigma(r_{M-1})},\dots,p_{\sigma(r_1)})$ ($r_{\gamma} \notin \{r_1,\dots,r_{M-1},r_{\alpha},r_{\beta}\}$) with the opposite signs (Fig.~\ref{fig:QQdag}). 
%$(-1)^{\frac{1}{2}|r_{\gamma}-r_{\alpha}| - |\lfloor \gamma-\alpha+\frac{1}{2} \rfloor|}$ and 
%$(-1)^{\frac{1}{2}|r_{\gamma}-r_{\beta}| - |\lfloor \gamma-\beta+\frac{1}{2} \rfloor|}$. 
At the same time, the operator $\tilde{\mathsf{Q}} \tilde{\mathsf{Q}}^{\dag}$ maps these restricted paths to each other with the same signs (Fig.~\ref{fig:QQdag2}). 
%$(-1)^{\frac{1}{2}(r_{\alpha}-r_{\beta})-(\alpha-\beta)}$ and 
%$(-1)^{\frac{1}{2}(r_{\beta}-r_{\alpha})-(\alpha-\beta)}$. 
Therefore, a linear combination 
$\sum c_{r_1,\dots,r_M} p^{\rm res}_{n}(p_{r_M}, \dots, p_{r_1})$, 
where the summation is taken for $r_1, \dots, r_M \in \{1,3,\dots,n-1\}$ ($r_1 < \dots < r_M$), 
is invariant under the anticommutator of the supercharges if the coefficients are chosen as 
\begin{equation} \label{eq:co_cond1}
\begin{split}
  c_{r_1,\dots,r_M} = 
  \begin{cases}
   1 
   & \{r_1,\dots,r_M\} = \{r'_2,\dots,r'_{M+1}\} \\
   (-1)^{-\frac{1}{2}(r'_j-r'_1) + (j-1)} & \{r_1,\dots,r_M\} = \{r'_1,\dots,r'_{M+1}\} \backslash \{r'_j\} \\
   0 & \text{otherwise} 
  \end{cases}
\end{split}
\end{equation}
for an arbitrary sequence $(r'_1,\dots,r'_{M+1})$ ($r'_j \in \{1,3,\dots,n-1\}$). 

\begin{figure}
 \centering
 \includegraphics[scale=0.55]{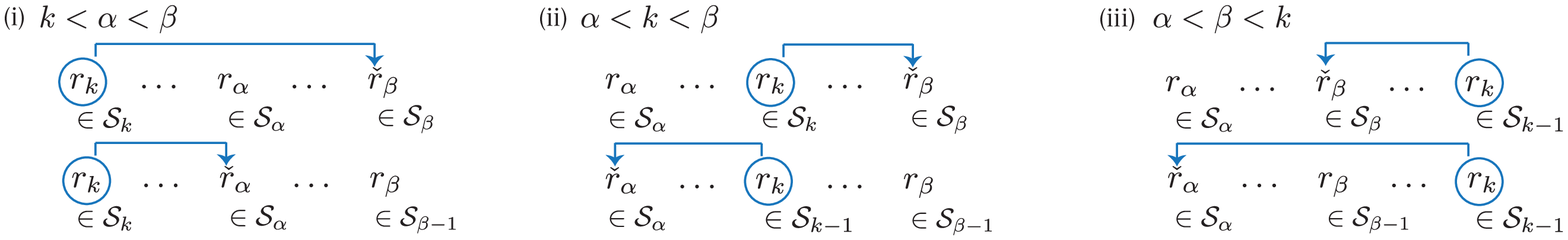}
 \caption{The possible configurations of $r_{\alpha},\,r_{\beta},\, r_{k}$ and the sets they belong to. In all cases (i)-(iii), the upper one represents a map $\tilde{\mathsf{Q}}^{\dag} \tilde{\mathsf{Q}}$ replacing $r_{k}$ with $\check{r}_{\beta}$, while the lower one represents a map replacing $r_{k}$ with $\check{r}_{\alpha}$. 
(i) The upper one gets $(-1)^{-\frac{1}{2}(r_k-\check{r}_{\beta}) - (k-\beta+1) - \frac{1}{2} (r_k-r_1) + \frac{1}{2}(\check{r}_{\alpha}-1) + (\alpha-1)}$ and the lower gets $(-1)^{-\frac{1}{2}(r_k-\check{r}_{\alpha}) - (k-\alpha+1) - \frac{1}{2} (r_k-r_1) + \frac{1}{2}(\check{r}_{\beta}-1) + (\beta-2)}$. 
(ii) The upper one gets $(-1)^{-\frac{1}{2}(r_k-\check{r}_{\beta}) - (k-\beta+1) - \frac{1}{2} (r_k-r_1) + \frac{1}{2}(\check{r}_{\alpha}-1) + (\alpha-1)}$ and the lower one gets $(-1)^{\frac{1}{2}(r_k-\check{r}_{\alpha}) + (k-\alpha-1) - \frac{1}{2} (r_k-r_1) + \frac{1}{2}(\check{r}_{\beta}-1) + (\beta-2)}$. 
(iii) The upper one gets $(-1)^{\frac{1}{2}(r_k-\check{r}_{\beta}) + (k-\beta-1) - \frac{1}{2} (r_k-r_1) + \frac{1}{2}(\check{r}_{\alpha}-1) + (\alpha-1)}$ and the lower one gets $(-1)^{\frac{1}{2}(r_k-\check{r}_{\alpha}) + (k-\alpha-1) - \frac{1}{2} (r_k-r_1) + \frac{1}{2}(\check{r}_{\beta}-1) + (\beta-2)}$. Thus we obtain that, in all cases, the upper one and the lower one result in the opposite signs. } \label{fig:QdagQ}
\end{figure}

\begin{figure}
 \centering
 \includegraphics[scale=0.55]{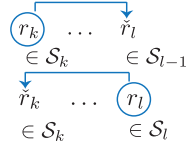}
 \caption{Under the map $\tilde{\mathsf{Q}}^{\dag} \tilde{\mathsf{Q}}$, the upper one gets $(-1)^{-\frac{1}{2}(r_k-\check{r}_l) - (k-l+1) - \frac{1}{2}(r_k-r_1) + (k-1) + \frac{1}{2}(\check{r}_l-1) + (l-2)}$ and the lower one gets $(-1)^{\frac{1}{2}(r_l-\check{r}_k) + (l-k-1) - \frac{1}{2}(r_l-r_1) + (l-2) + \frac{1}{2}(\check{r}_k-1) + (k-1)}$, which result in the same signs. } \label{fig:QdagQ2}
\end{figure}

The invariant linear combinations under the action of $\tilde{\mathsf{Q}}^{\dag} \tilde{\mathsf{Q}}$ are obtained as the superpartners of those invariant under $\tilde{\mathsf{Q}} \tilde{\mathsf{Q}}^{\dag}$. The explicit calculation leads to the action of $\tilde{\mathsf{Q}}^{\dag} \tilde{\mathsf{Q}}$ on a restricted path, which just changes the signs in the right-hand side of \eqref{eq:QdagQ} except for the first term. Suppose the $\tilde{\mathsf{Q}} \tilde{\mathsf{Q}}^{\dag}$-invariant restricted path of $M-1$ spinons 
$\sum c_{r_1,\dots,r_{M-1}} p_n^{\rm res}(p_{r_{M-1}},\dots,p_{r_1})$. 
Then this path is mapped by $\tilde{\mathsf{Q}}$ as 
\begin{equation} \label{eq:map_Q}
\begin{split}
 &\tilde{\mathsf{Q}}:
 \sum c_{r_1,\dots,r_{M-1}} p_n^{\rm res}(p_{r_{M-1}},\dots,p_{r_1})
 \\
 &\mapsto
 \sum_{j=1}^{M-1} \sum_{\check{r}_t \in \mathcal{S}_j} (-1)^{\frac{1}{2}(\check{r}_t-1)+(j-1)} c_{r_1,\dots,r_{M-1}} p_n^{\rm res}(p_{\sigma(\check{r}_t)},p_{\sigma(r_{M-1})},\dots,p_{\sigma(r_1)}). 
\end{split}
\end{equation}
This linear combination can be checked to be invariant under $\tilde{\mathsf{Q}}^{\dag} \tilde{\mathsf{Q}}$. Let us consider a sequence $(r_1,\dots,r_{M-1})$ ($r_j \in \{r'_1,\dots,r'_M\}$). According to \eqref{eq:map_Q}, a restricted path with $r_1,\dots,r_{M-1}$ is mapped to the one with $r_1,\dots,\check{r}_k,\dots,r_{M-1},r_{\alpha}$ ($r_{\alpha} \in \{1,3,\dots,n-1\} \backslash \{r_1,\dots,r_{M-1}\}$). For $r_{\alpha}, r_{\beta} \notin \{r'_1,\dots,r'_M\}$, two restricted paths in the combination \eqref{eq:map_Q} {\it e.g.} 
$p_n^{\rm res}(p_{\sigma(r_{\alpha})},p_{\sigma(r_{M-1})},\dots,p_{\sigma(r_1)})$ and 
$p_n^{\rm res}(p_{\sigma(r_{\beta})},p_{\sigma(r_{M-1})},\dots,p_{\sigma(r_1)})$ 
are mapped by the product $\tilde{\mathsf{Q}}^{\dag} \tilde{\mathsf{Q}}$ to the same restricted path 
$p_n^{\rm res}(p_{\sigma(r_{\beta})},p_{\sigma(r_{\alpha})},p_{\sigma(r_{M-1})},\dots,\check{p}_{\sigma(r_k)},\dots,p_{\sigma(r_1)})$ 
with different signs (Fig.~\ref{fig:QdagQ}). 
On the other hand, for $r_{\alpha} \notin \{r'_1,\dots,r'_M\}$ and $r_{\beta} \in \{r'_1,\dots,r'_M\} \backslash \{r_1,\dots,r_{M-1}\}$, a restricted path with $r_1,\dots,\check{r}_k,\dots,r_{M-1},r_{\alpha},r_{\beta}$ is obtained through the four different kinds of restricted paths: 
\begin{enumerate}
 \item[(i)] $p_n^{\rm res}(p_{\sigma(r_{\alpha})},p_{\sigma(r_{M-1})},\dots,p_{\sigma(r_1)})$ \qquad
	    $r_k$ replaced by $r_{\beta}$. 
 \item[(ii)] $p_n^{\rm res}(p_{\sigma(r_{\beta})},p_{\sigma(r_{M-1})},\dots,p_{\sigma(r_1)})$ \qquad
	     $r_k$ replaced by $r_{\alpha}$. 
 \item[(iii)] $p_n^{\rm res}(p_{\sigma(r_{\beta})},p_{\sigma(\alpha)},p_{\sigma(r_{M-1})},\dots,\check{p}_{\sigma(r_k)},\dots,p_{\sigma(r_1)})$ \qquad
	      mapped to itself. 
 \item[(iv)] $p_n^{\rm res}(p_{\sigma(r_{\beta})},p_{\sigma(\alpha)},p_{\sigma(r_{M-1})},\dots,\check{p}_{\sigma(r_{l \neq k})},\dots,p_{\sigma(r_1)})$ \qquad
	     $r_l$ replaced by $r_k$. 
\end{enumerate}
The cases (i) and (ii) always get opposite signs and therefore cancel each other out. The case (iv) has the non-zero contribution (Fig.~\ref{fig:QdagQ2}). 
%which is the $\tilde{\mathsf{Q}}^{\dag} \tilde{\mathsf{Q}}$-invariant restricted path. 
Thus, a linear combination 
$\sum c'_{r_1,\dots,r_M} p^{\rm res}_{n}(p_{r_M}, \dots, p_{r_1})$ 
is invariant if the coefficients satisfy 
\begin{equation} \label{eq:co_cond2}
 c'_{\sigma(r_1),\dots\sigma(r_{M-1}),\sigma(\check{r}_t)}
  = (-1)^{\frac{1}{2}(\check{r}_t-1)+(j-1)} c_{r_1,\dots,r_{M-1}}, 
\end{equation}
where $c_{r_1,\dots,r_{M-1}}$ meets the condition \eqref{eq:co_cond1}. 
%A linear combination of restricted paths 
%$\sum c'_{r_1,\dots,r_M} p^{\rm res}_{n}(p_{r_M}, \dots, p_{r_1})$ 
%is invariant under $\tilde{\mathsf{Q}}^{\dag} \tilde{\mathsf{Q}}$ if the coefficients are given by 
%\begin{equation} \label{eq:co_cond2}
%  c'_{r_1,\dots,r_M} = 
%  \begin{cases}
%   (-1)^{j-1} c_{r'_1,\dots,r'_{M-1}} &
%   \{r_1,\dots,r_M\} \backslash r_j = \{r'_1,\dots,r'_{M-1}\} \\
%   0 & \text{otherwise}. 
%  \end{cases}
%\end{equation}
%\begin{equation} \label{eq:co_cond2}
% c'_{r'_1,\dots,r'_M} = 
%  \begin{cases}  
%   c_{r_1,\dots,r_{M+1}} (-1)^{\sum_{k=1}^j \frac{r_k-r_{k-1}}{2}} 
%   & r_0 \equiv 1,\, r_j \equiv \{r_1,\dots,r_{M+1}\} \backslash \{r'_1,\dots,r'_M\} \\
%   0 & \text{otherwise}, 
%  \end{cases}
%\end{equation}
%where $c_{r_1,\dots,r_{M+1}}$ satisfies the conditions \eqref{eq:co_cond1}. 
%The conditions \eqref{eq:co_cond2} are obtained directly by applying $\tilde{\mathsf{Q}}$ to the linear combination given by \eqref{eq:co_cond1}. 
%\begin{equation} \label{eq:co_cond2}
%  \sum_{r_1,\dots,r_M \in \{s'_1,\dots,s'_{\frac{n}{2} - M + 1}\} \atop r_1<\cdots<r_M} \sigma'_{r_1,\dots,r_M} c_{r_1,\dots,r_M} = 0, 
%\end{equation}
%where $s'_1,\dots,s'_{\frac{n}{2} - M + 1}$ is randomly chosen from the set $\{1,3,\dots,n-1\}$. 
%$\sigma'_{r_1,\dots,r_M}$ is determined in such a way that satisfy 
%\begin{equation}
% \sum_{r_1,\dots,r_M} \sigma_{r_1,\dots,r_M} \sigma'_{r_1,\dots,r_M} = 0. 
%\end{equation}

It is checked that the linear combinations specified by \eqref{eq:co_cond2} are orthogonal to those given by \eqref{eq:co_cond1}. This verifies that the paths satisfying \eqref{eq:co_cond1} and \eqref{eq:co_cond2} make a superpartner. Indeed, the combination given by \eqref{eq:co_cond1} is killed by $\tilde{\mathsf{Q}}^{\dag}$, whereas that specified by \eqref{eq:co_cond2} is annihilated by $\tilde{\mathsf{Q}}$. Thus, the paths encoded by \eqref{eq:co_cond1} and \eqref{eq:co_cond2} span orthogonal eigenspaces of the supersymmetric Hamiltonian. Their eigenvalues are calculated as $\frac{n}{2}$ by using the properties obtained in Fig.~\ref{fig:QQdag2} and \ref{fig:QdagQ2}, which coincide with the exact results obtained from the spin chain Hamiltonian \eqref{eq:Hamiltonian_q0}.

\section{Conclusions}
In this paper, we discussed the discrete analog of the supersymmetry on the spinon basis of the infinite spin chain of the Fateev-Zamolodchikov (FZ) type at $q \to 0$. By modifying the supercharges introduced to the finite spin chain in order to be compatible with the infinite system, we showed that the actions on the spin basis properly realize the matrix forms and the comultiplications of the superalgebra on the supersymmetric sine-Gordon model. We also constructed the linear combinations of paths which constitute the supersymmetric subspace of the FZ chain.  
%Thus, our aim, {\it i.e.} to investigate the supersymmetry on the spinon basis, was achieved for the Fateev-Zamolodchikov spin chain. 
We found that, although the discrete analog of the supercharge inserts or removes a site in the spin basis, it preserves the number of spinons and the spin of each spinon. Among two kinds of degrees of freedom, the $SU(2)$ degrees of freedom and the RSOS degrees of freedom, the supercharge makes a non-trivial action only on the latter. 

Since our motivation come from the lattice regularization of quantum field theories, which can be partially achieved through the light-cone lattice regularization~\cite{bib:ABPR08, bib:ANS07, bib:FRT98, bib:M14}, we are interested in how the discussion here works for finite spin chains. The problem to overcome is that, since the vertex operator is well-defined only in infinite systems, the spinon creation operators used in this paper cannot be used in a finite chain. 
One possibility is to use the hole description of spinons via the Bethe ansatz method~\cite{bib:BKMW04}. In the Bethe ansatz method, each eigenstates is characterized by a set of Bethe roots, which are classified into strings~\cite{bib:EKS92, bib:FM01, bib:TS72}. From the string compositions of eigenvectors, the Yangian-like structure was found for the spin-$\frac{1}{2}$ case~\cite{bib:AKMW06}. The similar structure is expected for the arbitrary spin-$\frac{\ell}{2}$ case, by letting configurations of shorter strings than length two give the RSOS degrees of freedom~\cite{bib:R91}. 
For the spin-$\frac{1}{2}$ case, the spinon orbitals were identified in association with spinon momenta~\cite{bib:AKMW06}. However, the method seems to be empirical and therefore, more sophisticated formulation would be needed for the extension to the spin $1$, which contains more degrees of freedom. 

Another future problem is to consider the arbitrary spin cases. Although we used the spinon creation operator defined at the $q \to 0$ limit, in which the eigenvectors admit the domain-wall description, the spin chain with spin $\frac{\ell}{2}$ does not generally possess the supersymmetry at arbitrary anisotropy but is supersymmetric only at the combinatorial point $q = e^{\frac{\pi i}{\ell+2}}$~\cite{bib:H13}. 

The problem to work on finite systems is especially interesting for us, since the FZ chain with boundary magnetic fields shows the phase transition in its correspondence with the supersymmetric sine-Gordon model~\cite{bib:M14}. That is, the spin chain realizes either the Neveu-Schwarz sector or the Ramond sector of supersymmetry depending on the boundary parameters. We expect that the study of spinon excitations and length-change operators may lead to understanding how the boundary condition of spin chains works for the hidden supersymmetry of spin chains, and subsequently, how the supersymmetry holds/changes in manipulating the scaling limit.

\section*{Acknowledgements}
The author thanks M. Staudacher and J. Suzuki for fruitful discussion about the discrete analog of the supersymmetry. The author is also grateful to K. Schoutens for inspiring interesting future works. This work is supported by CREST-JST and JSPS Grant-in-Aid No.~15K20939.

%Bibliography
\bibliographystyle{abbrv}
\bibliography{references}

\end{document}